%% file: SW1.tex
\documentstyle{europhys}
\input euromacr

\begin{document}
\euro{xx}{y}{1-6}{1998}
\Date{13 August 1998}
\shorttitle{P. van der Schoot \etal LINEAR AGGREGATION REVISITED ETC.}
\title{Linear Aggregation Revisited: Rods, Rings and Worms}
\author{Paul van der Schoot\inst{1}\thanks{E-mail: vdschoot@mpikg-teltow.mpg.de} \, 
and Joachim P. Wittmer\inst{2}\thanks{E-mail: j.wittmer@ed.ac.uk} 
}
\institute{
\inst{1} Max-Planck-Institut f\"{u}r Kolloid- und Grenzfl\"{a}chenforschung\\ 
         Kantstrasse 55, D-14513 Teltow-Seehof, Germany\\
\inst{2} Department of Physics and Astronomy, University of Edinburgh\\
              Mayfield Road, Edinburgh EH9 3JZ, UK }
\rec{20 August 1998}{}
\maketitle
%
%
%


\begin{abstract} 
The problem of ring formation in solutions of 
cylindrical micelles is reinvestigated theoretically, taking into account 
a finite bending rigidity of the self-assembled linear objects. Transitions
between three regimes are found when the scission energy is sufficiently large. 
At very low densities 
only spherical and very short, rod-like micelles form. 
Beyond a critical density, mainly rings but also rod-like chains appear in  
(virtually)
fixed relative amounts.  
Above a second transition both the length of the linear chains and 
the relative amount of material taken up by them
increase rapidly with increasing concentration.
The mass accumulated into long, semi-flexible
worms then
overwhelms that in rings. 
The ring-dominated regime is very narrow for semi-flexible chains, 
confirming that the presence of 
rings may be difficult to observe
in many micellar systems, and indeed disappears completely for 
sufficiently low scission
energy and/or large persistence length.
\end{abstract}

Solutions of highly elongated, cylindrical micelles are 
arguably among the best studied 
of the so-called equilibrium polymeric systems \cite{CatesCandau}. 
Equilibrium polymers
are formed in a reversible polymerisation process and therefore are in 
chemical equilibrium with each other -- monomeric material is 
continually exchanged between the assemblies. An aspect not at all well
understood is why ring closure seems to be
unimportant in solutions of linear micelles,
although this would remove 
unfavourable free ends (``end caps") from the solution. (Closed loops have been 
observed in 
electron microscopic
images of linear micelles \cite{Clausen}, but apparently 
they occur in too low concentrations 
 to significantly influence the properties of 
 micellar systems \cite{CatesCandau}.)
In other equilibrium polymeric systems, such as liquid sulphur, the presence of
 rings is on the other hand thought to be 
all-important \cite{PPW}.
According to mean-field theory \cite{PPW}, 
rings must indeed
overwhelmingly
dominate the aggregate population 
in solutions of self-assembled,  {\it flexible}  polymers, but only when the monomer 
 fugacity is below 
some threshold
value. At fugacities above this threshold, 
the concentration of rings should level off to a constant value. Linear chains then
form an
increasingly important fraction of the aggregated material. 
In the limit where the end caps carry an infinite free energy penalty, 
the crossover becomes sharp and is reminiscent of a phase transition, sometimes 
referred to as the polymerisation transition \cite{PPW}.

Ignoring 
the subtle effects of excluded volume interactions which give rise to departure
from mean-field
behaviour \cite{PPW, MEC, Schoot, Wittmer}, 
the self-assembled linear chains behave as if the rings were not present 
above the polymerisation 
transition,
although the rings
do deplete a fixed amount of
 material available to linear aggregation \cite{Porte}. The 
amount of material incorporated into rings is then roughly
equal to the volume fraction at the polymerisation threshold which, for flexible aggregates,
is predicted to 
take place at an estimated value of the order ten per cent. This
 is at variance with 
experimental observations on micellar solutions, where in some
cases giant, polymer-like assemblies are known
to arise at volume fractions as low as one tenth 
of a per cent \cite{CatesCandau}.
Some workers
 have
conjectured that {\it a finite bending rigidity} of the micelles could suppress rings smaller than
a certain size, thereby reducing the contribution of rings \cite{MEC}. Others
argued that the formation of rings is anyway more strongly attenuated than 
any estimate based on
a Gaussian bead-spring model can predict. 
This is essentially because the number of places
a continuous ring can break is much greater than that of a discrete ring \cite{Porte}.
Although both views are
not entirely inaccurate, we shall see below that the issue is decidedly more
complex and worth of study.

In this Letter, we reinvestigate linear self-assembly in three spatial dimensions
by explicitly taking into account
the finite bending rigidity of the chains. We find 
three regimes, one where spherical and possibly also short, rod-like
aggregates are overwhelmingly present, one where closed rings constitute
a significant if not dominant portion of the micelle
population, and one where long, worm-like 
assemblies emerge to become the dominant species,
see figure 1.
For semi-flexible linear assemblies containing many monomers per persistence length,
the crossovers between the various regimes occur at such low volume fractions that
for most practical purposes ring formation can be ignored. This explains
experimental observation, and is in line
with an earlier (less elaborate) analysis of Porte \cite{Porte}. We also find
conditions under which the ring phase disappears altogether.

\begin{figure}
\includegraphics{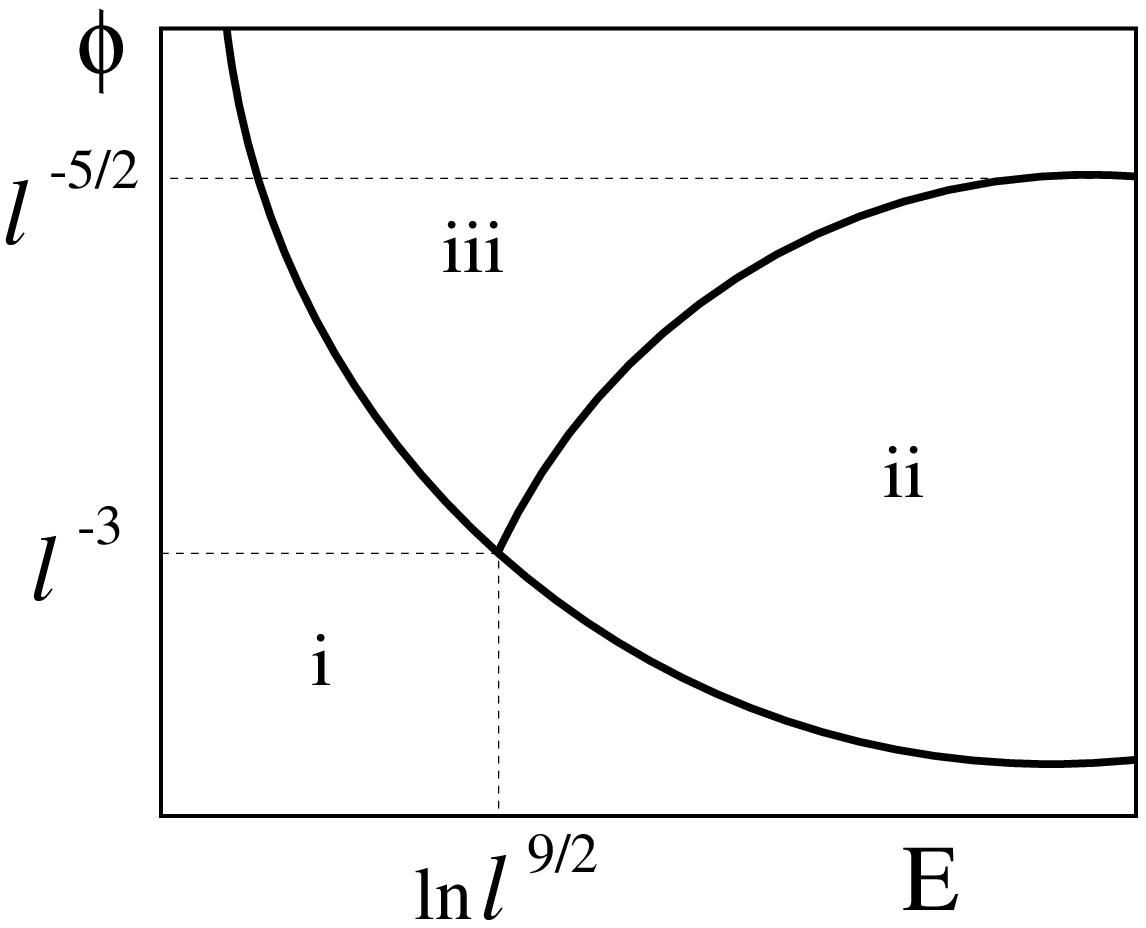}
\includegraphics{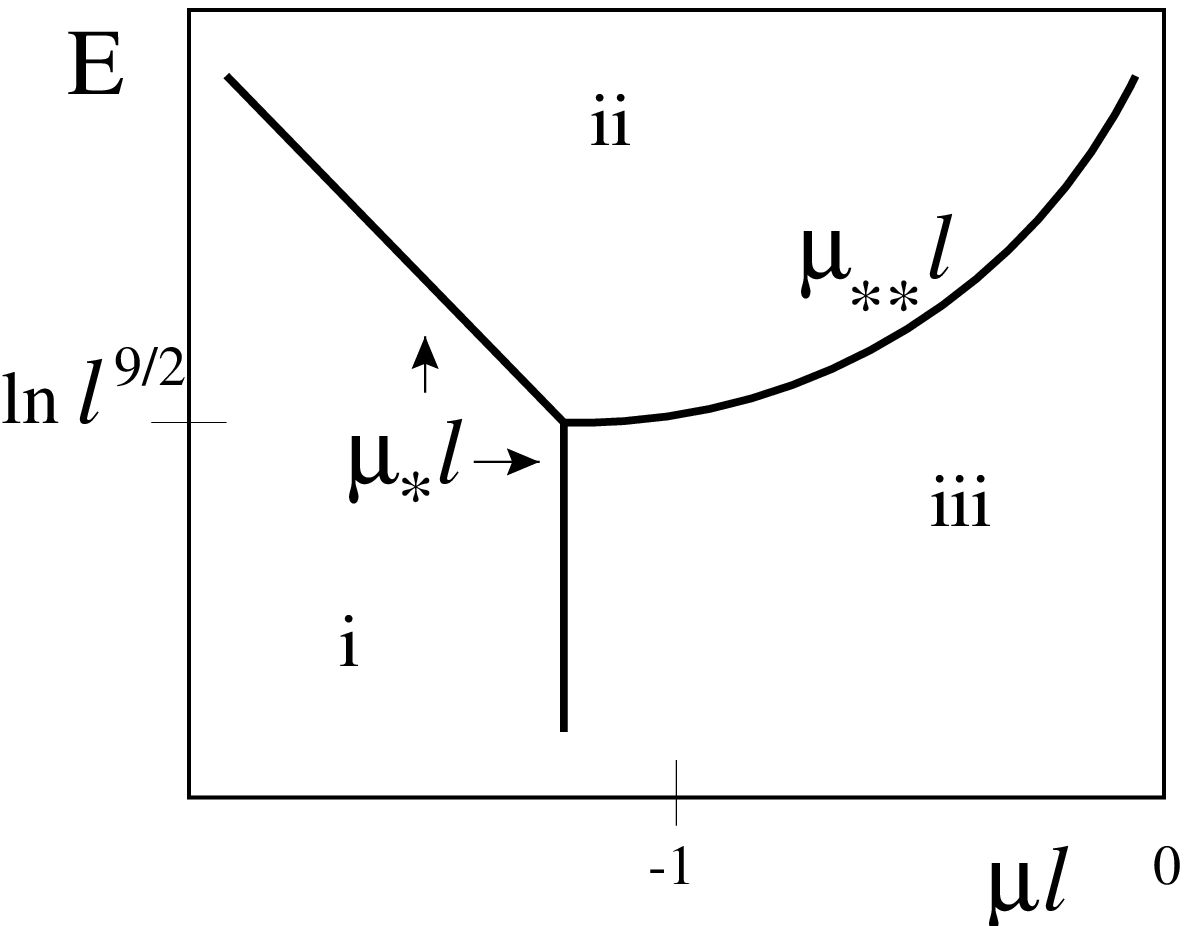}
\vskip 4.5cm
\caption{Schematic diagram of the aggregation states of semi-flexible micelles. 
Volume fraction $\phi$ {\it vs} the scission energy $E$. 
(i) Monomeric phase (spherical micelles), 
(ii) ring phase consisting mainly of rings and a few 
rods, both about a persistence length long,  
(iii) worm phase of long, worm-like aggregates, many persistence lengths longs, 
and a few relatively short rings. 
Here $l$ is a dimensionless persistence length and unimportant prefactors of 
order unity have been left out.  \label{fig1}}
\caption{As figure 1, now given as a function of the scission energy $E$, 
and the chemical potential $\mu$ times the dimensionless persistence length $l$. 
Indicated are also the cross-over chemical potentials $\mu_*$ and
$\mu_{**}$, discussed in the main text.
\label{fig2}}
\end{figure}

We consider a micellar solution above the critical micelle concentration, cmc.  
We assume that at (or slightly above) the cmc the micelles are essentially spherical, and that
we can disregard
the presumably constant amount of the dissolved surfactant not incorporated into micelles.
In our model, the spherical micelles are viewed as the basic monomeric units that 
self-assemble into elongated, continuous chains. Whether in actual fact
single surfactant molecules self-assemble into cylinders or whole groups
of surfactants is inessential to the theory. The cylindrical assemblies can be open 
(linear) or closed (circular), and are assumed to be semi-flexible. This implies that 
the length each ``monomer" contributes to the total
length is considerably smaller than the persistence length. Self-interactions of 
the excluded volume type are comparatively weak in solutions of semi-flexible 
chains \cite{Grosberg}, so the chains may be treated as ideal. The dimensionless number
density of assemblies of aggregation number $N$ is denoted $\rho (N)$, where
the subscripts $0$ and $1$ shall be used to refer to 
closed rings  and linear chains, respectively. Balancing 
the loss of translational entropy
of a monomer when incorporated into an aggregate against the free energy gain of 
association immediately gives, in the mean field,
$\rho (N) = Z(N) \exp (\tilde{\mu} N)$
for the equilibrium size distribution. The chain
partition function $Z(N>1)$ depends on the topology of the aggregate
as well as on the prevailing flexibility mechanism. 
We take as a reference the ``internal" partition
function of a free monomer, and define $Z(1) \equiv 1$. 
The (dimensionless) chemical potential of the monomers, $\tilde{\mu}$, can be related to the 
total
volume fraction of aggregating material,  $\phi$, via the normalisation condition 
$ \phi = \phi_{0} + 
\phi_{1}$, where $\phi_0  \equiv \sum_{N=2}^{\infty}   N \, \rho_{0} (N) $ and $\phi_1
\equiv \sum_{N=1}^{\infty}  
N \,\rho_{1} (N) $. In our calculations below, we replace summations 
by integrations 
whenever the mean aggregation numbers 
are large, that is, whenever $\left< N \right>_0 
\equiv \phi_0 / \sum_{N=2}^\infty 
 \rho_0 (N)
 \gg 1$ and/or $\left< N \right>_1 \equiv \phi_1 / \sum_{N=1}^\infty 
 \rho_1 (N)
\gg 1$. 

Given that the size distribution is directly proportional to the partition
function $Z$, one would next like to obtain accurate expressions for the 
partition functions $Z_{0}$ and $Z_{1}$ for arbitrary aggregation numbers $N$. 
Within the rigid Gaussian chain model of semi-flexible linear chains,
discussed recently by Marques and Fredrickson \cite{Marques}, 
the latter can 
be evaluated exactly in the entire range from the rod to the worm limit. We 
only quote the limiting results for the rod and worm limits, which can
be expressed as
$Z_{1} (N) \sim z^{N-1}  ((N-1)/l)^{3(N-2)/2}$ and
$Z_{1} (N) \sim z^{N-1} (2l)^{-3/2} $ in these two limits $l \gg
N \geq 2$ and $ N \gg l $. Here 
$l \gg 1$ denotes the average
number of monomers per persistence length, presumed large, and $z  \equiv \lambda_1
l^{-3/2}  \exp (\tilde{E})$ is a function of $l$ and $\tilde{E}$, with $\lambda_1$ a constant of order unity. 
A Boltzmann term associated with the (dimensionless) free energy cost
 of chain scission, $\tilde{E} > 0$, has been explicitly added, penalising
the free ends of the open chains. For long, polymer-like micelles values of $\tilde{E}$ 
 up to about 20 ($k_B T$) are not uncommon \cite{CatesCandau}, although scission energies as high
as 38 ($k_B T$) have been reported in the literature \cite{Zana}. Note i) that the
free energy of a single chain, $ - \ln Z$, is extensive in the number of monomers $N$
in the chain, as it should, but  ii) that there are logarithmic corrections 
near the
rod limit. These corrections very strongly suppress 
rod-like aggregates considerably smaller than one persistence length yet consisting
of more then a few aggregating units. (We shall come back to
this later.) As $Z_1(N)  \ll z^{N-1} (2l)^{-3/2}$ for $2 \ll N \ll l$, we simply
set $Z_{1} (N) = 0$ for the entire range $2 \leq N < l$, and $Z_1 (N) = z^{N-1}
(2l)^{-3/2}$
for $N \geq l$
. The free monomers are from now on understood to 
also include short, rod-like micelles a few micelle diameters long. This merely renormalises the 
chemical potential.

Exact results for
the partition function of closed
loops of arbitrary flexibility are, as far as we are aware, not available 
in the literature, although 
limiting expressions have been obtained
within several model descriptions. 
The problem of ring closure in the tight bending limit was addressed by
Shimada and Yamakawa at the level of the worm-like chain model \cite{Yamakawa}.
Their result can be
written as $ 
Z_{0} \propto \exp ( - 2 \pi^2 l/N )$, valid when $l \gg N \gg 1$. (Here we ignore
additional
non-exponential
dependences on $N/l$ for reasons of brevity.) In other words, tight
rings are exponentially suppressed. Again, we  
arbitrarily set the cut-off at $N = l$, and put $Z_{0} = 0$ for all $1 \leq N < l$. 
(There
is obviously no monomeric ring.) In the opposite
limit of long chains, $N \gg l \gg 1$,
a simple scaling argument allows us to deduce 
the partition function of the closed configurations, $Z_{0}$, 
from that of the linear chains, $Z_{1}$.  
The argument is a slight modification of the one given by Porte \cite{Porte} 
(and others \cite{PPW, MEC}) for the case of
flexible chains, and hinges on a 
consideration of the ratio $Z_{1} / Z_{0}$.
The ratio $Z_{1} / Z_{0}$ is equal to the probability
of opening a loop, which
must be proportional to i) the number of places the ring can break, $N$,  ii) 
the volume that two neighbouring segments can explore after being disconnected, 
proportional to
$N^{3/2} l^{3/2}$ \cite{Porte},
 iii) the increased angular phase 
volume of these new end segments, $l^{1/2}$,
and iv) a Boltzmann weight, $\exp (\tilde{E})$, for the 
creation of the two end caps. Using the previously given expression for $Z_1$,
we find $Z_{0} \sim \lambda_0 
z^{N} N^{-5/2} l^{-2}$
for $N \gg l \gg 1$, with $\lambda_0$ an unknown constant presumably close to
unity.  (Pfeuty and co-workers \cite{PPW} set $\lambda_{0} = 0.1$ in their
analysis of self-assembled,  flexible chains.) 
We extrapolate this expression in our calculations down to $N = l$.
Our estimate for $Z_{0}(N > l)$ is smaller by a factor of
$l^{-1/2}$ than the previous estimate of Ref.\ \cite{Porte}, 
due to the reduced angular phase volume of the two ends upon ring closure, 
not taken into account in that work.

The size distributions we thus obtain  are
\begin{equation}
\rho_{0} (N) \simeq \left\{ \begin{array}{ll} 0 &
\mbox{if $1 \leq N < l$} \\ \lambda_0 N^{-5/2} l^{-2} \exp ( \mu N) & \mbox{if $N \geq l$} \end{array} \right.
\end{equation}
for the rings, and
\begin{equation}
\rho_{1} (N) \simeq \left\{ \begin{array}{ll} z^{-1} \exp (\mu ) & \mbox{if $N = 1$} \\
0 &
\mbox{if $2 \leq N < l$} \\ \exp ( -E + \mu N) & \mbox{if $N  \geq l$} \end{array} \right.
\end{equation}
for the linear chains, where in both expressions we have absorbed a $\ln z$ in the
chemical potential, $\mu \equiv \tilde{\mu} + \ln z$, and a constant into the scission
energy, $E \equiv \tilde{E} - \ln (2^{3/2}/\lambda_1 )$, which 
we regard as the
actual, measurable scission energy. Using eqs (1) and (2), we 
evaluate the relative contributions of chains and rings to the total volume fraction
material $\phi$, and calculate their mean lengths $\left< N\right>$. Careful inspection
of the various contributions to
\begin{equation}
\phi_0 \simeq 2 \lambda_0 l^{-5/2} \left( \exp (\mu l) - \sqrt{- \pi \mu l} \,  \mbox{erfc} (\sqrt{-\mu l})  \right)
\end{equation}
and to
\begin{equation}
\phi_1 \simeq z^{-1} \exp (\mu ) +  l^2  \exp (\mu l - E) \left( \frac{1}{(\mu l)^2}
- \frac{1}{\mu l} \right)
\end{equation}
reveals that, depending on the dimensionless chemical potential $\mu \leq 0$,
three regimes can be distinguished. See figures 1 and 2. When $E \gg \ln (l^{9/2} 
\lambda_0^{-1} ) $ we find, at a fixed scission energy $E$, transitions between all
three regimes upon increasing the chemical potential:

{\bf (i)} The {\bf monomer-dominated} regime: $\mu < \mu_*$, where $ \mu_* \simeq
- \, l^{-1} \ln \left( \lambda_1 l^{1/2} + 
\lambda_0 \lambda_1 l^{-4} \right. $ $\left. \times
\exp (E) \right)$. For volume fractions $\phi < \lambda_1^{-1} l^{3/2} 
\exp (-E)$, almost all of the material 
resides in ``monomers" with $\left< N \right>_1 = 1$, i.e., in spheres
and (possibly) in very short rods.

{\bf (ii)} The {\bf ring-dominated} regime: $\mu_* < \mu < \mu_{**}$, where $\mu_{**} \simeq
- l^{5/4} (2 \lambda_0)^{-1/2} \exp (-E/2) $. When $\mu_* < \mu < -l^{-1}$, 
rings and chains of about a persistence length long occur in a 
ratio almost independent of the concentration, $\phi_1 / \phi = 1 - (\phi_0 / \phi )
 \simeq 1/(1 + \lambda_0
l^{-9/2} \exp (E) )$. The mean lengths $\left< N \right>_0 \simeq \left< N \right>_1 
\simeq l \, (1-(\mu l)^{-1}) \gg 1$
increase only logarithmically with $\phi$. When $-l^{-1} < \mu < \mu_{**}$ 
the mass fraction of worms grows with increasing chemical
potential.

{\bf (iii)} The {\bf worm-dominated} regime: $\mu_{**} < \mu \leq 0$. The fraction of 
worms overwhelms the ring
fraction when $\mu > \mu_{**}$, that is, when
$\phi > 4\lambda_0 l^{-5/2}$. The volume fraction
of material held in rings levels off to $\phi_0 \simeq 2 \lambda_0 l^{-5/2}$, 
as does the mean size  $\left< N\right>_0 \simeq 3 l$. We  find, furthermore, 
that in this worm-dominated
regime
$\left< N\right>_1 = l \, (\phi - \phi_0)^{1/2} \exp (E/2) \gg l $. 
This is consistent with Porte's result \cite{Porte}, except for the different
scaling of $\phi_0$ with l.


When  $E < \ln (l^{9/2} \lambda_0^{-1} )$, in other words when
the micelles are sufficiently stiff for a fixed scission energy $E$,
the ring phase 
disappears and we are left with a single transition from a ``monomeric" solution 
of spherical and short, rod-like micelles to
a  worm-dominated regime. The 
cross-over takes place when $\mu > \mu_* \simeq - (\mu_* l)^{-1} \exp (- \mu_* l)$, 
which
for all sensible values of $l$ gives $- \mu_* l = O(1)$ independent of the scission
energy. The corresponding cross-over volume fraction, however, does depend on $E$, $\phi \simeq \lambda_1^{-1}
l^{3/2} \exp (-E) < \lambda_1^{-1} l^{-3} \ll 1$.
See also 
the ``phase" diagrams given
in figure 1 (in $\phi -E$ space) and in figure 2 (in $E-\mu$ space). Note that ``monomers" do occur in 
the polymerised phases (ii) and (iii),
albeit only in very low concentrations. In the electron micrographs
of Ref.\ \cite{Clausen}, spherical micelles can indeed be seen in 
coexistence with long, polymerlike micelles. 

One may infer from our calculations that the so-called sphere-to-rod transition 
could actually be a manifestation of the two transitions discussed above. This is plausible
as the
intermediate (ring) regime can be very narrow indeed for cylindrical assemblies 
which are quite rigid on the scale of a micelle diameter \cite{vBerlepsch}. 
Even when the assemblies are not
all that stiff, the ring regime may be difficult to detect as the
worm-dominated phase sets in at very low volume fractions of order $l^{-5/2}$. 
Indeed, for 
fairly flexible linear micelles, such as those formed by the surfactant
CPyBr in brine, we
(conservatively) estimate that $l \simeq 4$, meaning that 
the worm phase must occur at a volume fraction of the order of $10^{-3}$ 
above the cmc. (Here $l$ was equated to the ratio of the persistence length
and the micelle diameter, which for this system are accurately known quantities \cite{PorteReview}.) 
Obviously, we are stretching the theory 
to the limit of its validity in this example,
but it does once more confirm that rings should almost
always be neglected in micellar systems \cite{Porte}.

Perhaps the most surprising conclusion of our analysis is that not
only tight rings are suppressed, but also rod-like assemblies shorter
than, say, a persistence length but longer than a few micelle diameters.
The reason is that close to the rod limit 
very few bending modes remain that could, with the associated configurational entropy,
help to compensate for the loss of translational entropy upon aggregation.
It has to be noted that 
the {\it precise} impact of the freezing out of bending modes
on the size distribution appears to be non-universal, i.e. it is model dependent. 
Indeed, other workers \cite{Ha, Morse, Winkler} do not  
find the relevant logarithmic terms in the single chain free energy at all. 
(In \cite{Ha, Winkler} this is likely to be due to the 
approximate evaluation of $Z_1 (N)$, and to the highly unusual choice of 
chain Hamiltonian in \cite{Morse}.)
The question of how strongly our results depend on the 
choice of the rigid Gaussian chain model \cite{Marques} 
needs to be addressed. 
This particular  model exhibits ``breathing" modes  which are
not present in the standard persistent or worm-like chain model \cite{Grosberg}, 
and could therefore exhibit spurious behaviour near the rod limit. 
It is, however, straightforward to verify that  
terms not identical but similar to those found in the rigid Gaussian chain model
do arise in a continuous description of the 
(inextensible)  worm-like chain model, 
 when taking the limit $N \rightarrow \infty$, $l \rightarrow \infty$, 
$N/l \rightarrow 0$ \cite{Notshown}. (In this limit the problem can be
treated at the level of a Gaussian fluctuation theory around
the straight-rod configuration.)  
Again we find that
the rod-like configurations in $\rho_1 (N)$ are strongly suppressed, 
be it with a different prefactor
of $N^{-N}$ for $1 \ll N \ll l$. It is for this reason that we expect  
 the general features
of our ``phase" diagram to be universal, 
at least in the limit $l \gg 1$.

Computer simulation studies of self-assembled lattice chains with a 
finite bending energy 
have recently
appeared in the literature \cite{Milchev, Rouault, Carl}. Unfortunately, 
comparison with our predictions is not straightforward, for  the
chains interact strongly in the quoted works (and in fact exhibit order-disorder
transitions)
 in 
contrast to the dilute case studied here. 
 Ring formation is furthermore explicitly forbidden in \cite{Rouault, Carl}, 
where simulations were in addition restricted to
two spatial dimensions only. (Another difficulty might arise from
the rather low mean chain sizes of order
ten investigated 
 in \cite{Milchev, Rouault, Carl}.) Tentatively setting aside these reservations, 
some of the 
trends obtained from our analysis
seem to be preserved in the simulations. Indeed, in \cite{Milchev} rings were found
to play an insignificant role in the (``disordered") worm-dominated phase, while in
\cite{Rouault, Carl} an increasing chain rigidity strongly reduces the mean 
aggregation number \cite{Note2}. This happens, in our calculation, 
close to the (i) $\rightarrow$ (iii) transition \cite{Notshown}. 

In conclusion, we have have shown, using a simple mean-field analysis, that a 
finite chain rigidity may be the reason why ring formation is unimportant in
many (if not all) micellar systems. As  mean-field theory
is thought to have a very broad range of validity in semi-flexible systems
due to a combination
of weak self-interactions and a narrow scaling regime \cite{Grosberg}, our results should be reasonably
accurate all the way from the dilute to the concentrated regime, 
provided the system does not cross over to a liquid crystalline phase 
where a different behaviour has to be expected \cite{Milchev}.

\stars
We thank Peter Swain for a critical reading of the manuscript.
 


\end{document}

%% file: euromacr.tex
\def\etal{{\hbox{{\tenit\ et al.\/}\tenrm :\ }}}

\def\stars{\bigskip\centerline{***}\medskip}

\newif\ifboo \boofalse
